# Extreme Flaring of Starlink Satellites

Anthony Mallama[1]* and Richard E.Cole

2024 May 20

[1] IAU - Centre for the Protection of Dark and Quiet Skies from
Satellite Constellation Interference

* Correspondence: anthony.mallama@gmail.com

Abstract

Starlink satellites can become extremely bright when sunlight reflects specularly to an observer on the ground. The observed brightness of such flares is consistent with a bidirectional reflectance function of the Starlink satellite chassis. These findings are applied to the case of an extreme flare that was reported as an Unidentified Aerial Phenomena by the pilots of two commercial aircraft.

1. Introduction

The brightness of a spacecraft may surge when sunlight reflects from one of its flat surfaces to an observer. The amplitudes of such *flares* are typically one magnitude or less for Starlink satellites (Mallama 2021 and Mallama et al 2024). This paper reports on extreme flares of many magnitudes that occur when sunlight reflects off the nadir side of the chassis at a glancing angle.

Section 2 describes the geometry of extreme flares and presents observations that characterize their brightness. Section 3 applies that characterization to a case where airline pilots reported exceedingly bright flaring Starlink satellites as Unidentified Aerial Phenomena.

2. Geometry and observations of extreme flares

Starlink satellites have flat panel chassis whose sides are zenith-nadir facing with a reflective surface on at least a portion of their nadir side. SpaceX applies a dielectric coating in order to reduce the adverse impact of their satellites on astronomical observing by mitigating satellite brightness. The coating is most effective for the observer when spacecraft are overhead and the mirror-like surface reflects sunlight specularly into space rather than allowing it to reflect diffusely toward the ground as shown in Figure 1.

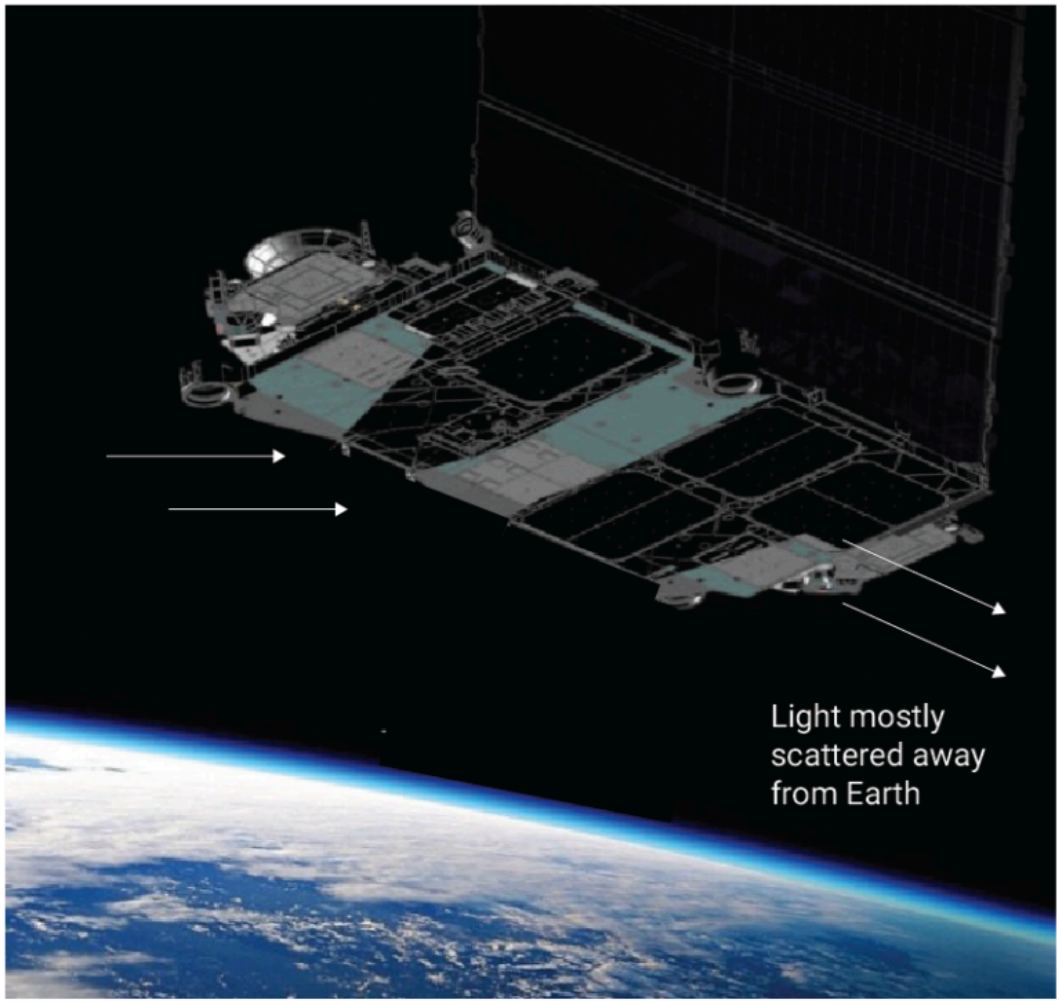

*Figure 1. Reflection of sunlight from the dielectric coating on Starlink version1.5. Credit: SpaceX.*

However, there is one scenario where satellites may appear brighter instead of fainter. Figure 2 illustrates the geometry where a Starlink satellite is seen near the horizon toward the solar azimuth. Sunlight is reflected to the observer in this scenario and the satellite may appear very bright.

Figure 2 also depicts the off-specular angle which is the arc measured at the satellite between the specular reflected ray and the direction to the observer. Geometrical analysis indicates that the off-specular angle is very small when a satellite is near the horizon and the Sun is at least 30º below the horizon.

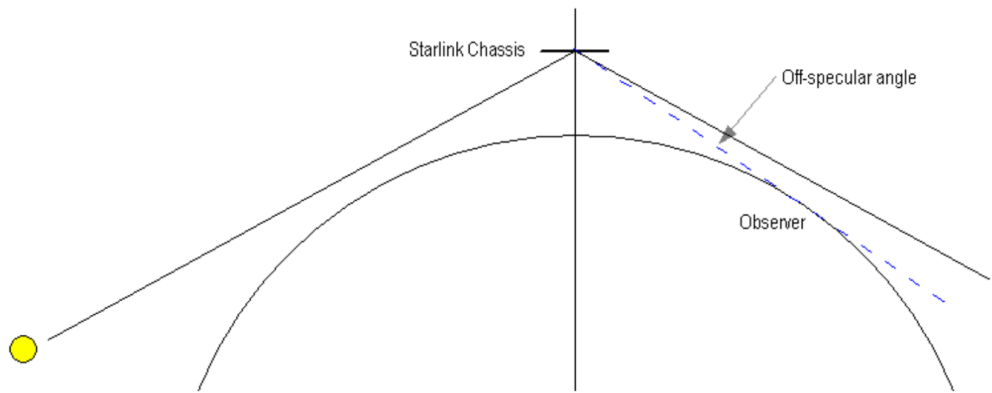

*Figure 2. Sunlight may be reflected toward the observer when a satellite is low above the horizon and toward the solar azimuth. The off-specular angle is described in the narrative.*

We investigated the relationship between satellite brightness and off-specular angle empirically. Visual magnitudes were obtained by author Mallama in the manner described by Mallama (2022). These observations targeted spacecraft near the solar azimuth over a range of satellite and solar elevations.

The strong dependence of satellite brightness on off-specular angle is illustrated in Figure 3. All four models of Starlink spacecraft can be fitted with log functions. The magnitudes on the left-hand side of the graph show that satellites can flare very brightly.

The empirical brightness function of Starlink version 1.5 satellites is especially interesting because SpaceX has provided the bidirectional reflectance distribution function (BRDF) for its chassis. Figure 4 shows the observed and BRDF functions at the same vertical axis scale, and the close correspondence validates both. The thinner lines represent the confidence limits of the best fit to the observations. These brightness functions are applied to the topical observation studied in the next section.

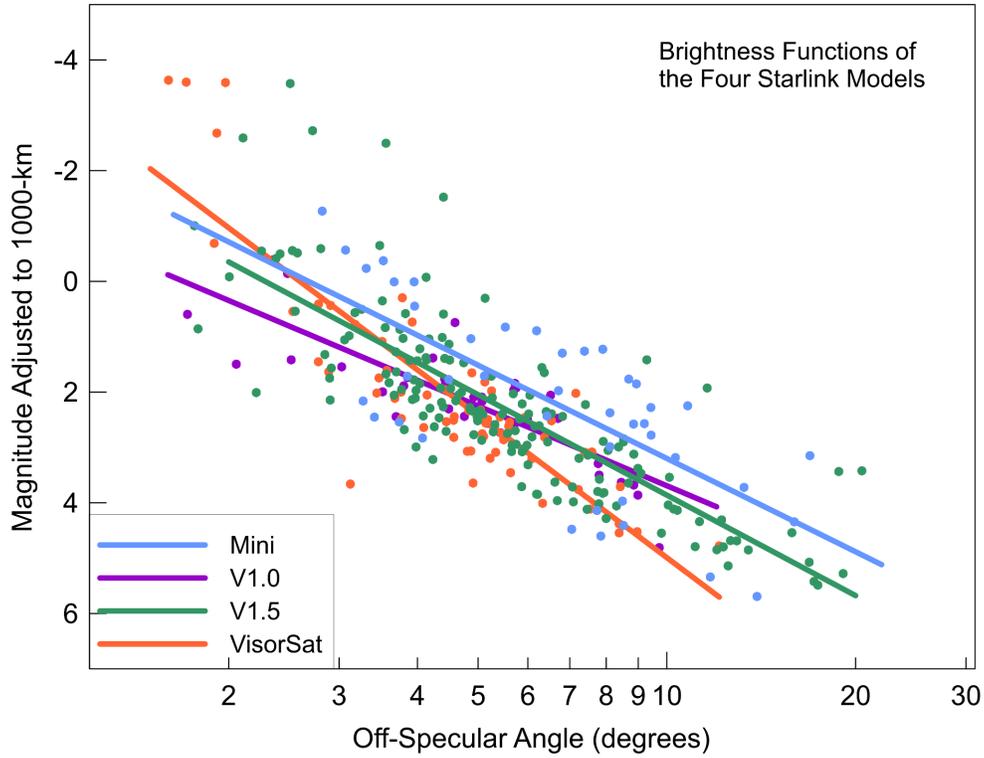

*Figure 3. Brightness versus off-specular angle for all models of Starlink.*

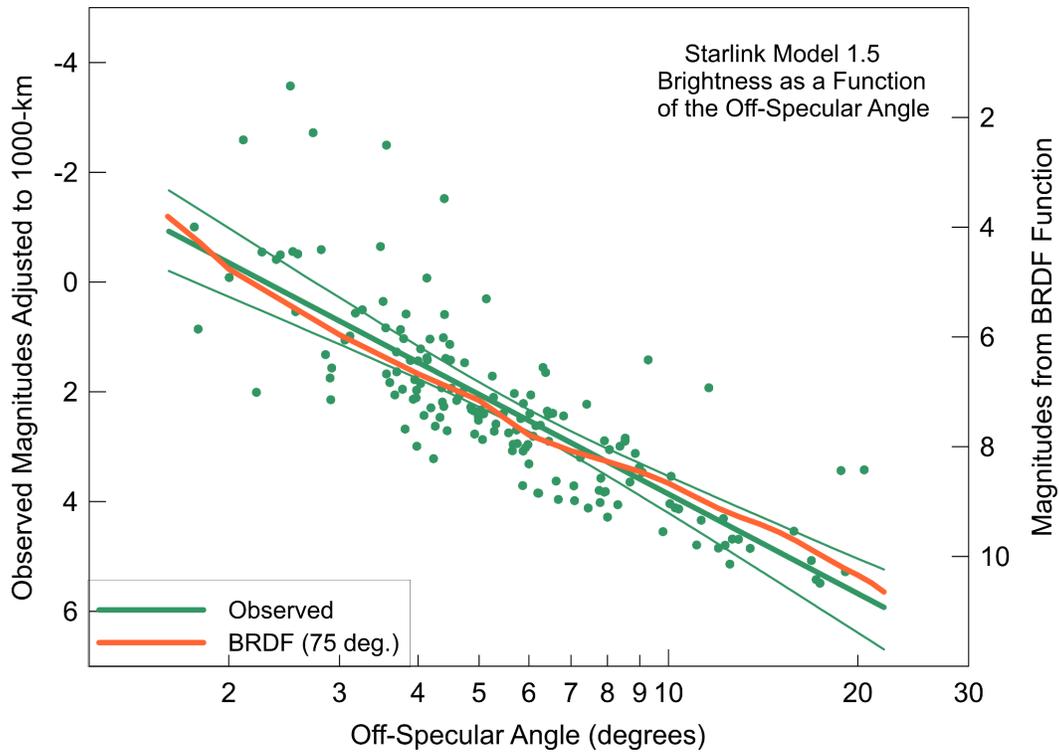

*Figure 4. Agreement between observed and BRDF functions for V1.5 Starlinks.*

3. Unidentified Aerial Phenomena

Airline pilots from two commercial flights witnessed a group of very bright objects moving across the sky and reported them as Unidentified Aerial Phenomena in 2022. Buettner et al (2024) identified those objects as a closely spaced train of Starlink satellites from launch 4-26 based upon the spacecrafts' positions at the time of the observation. The measurement of brightness from photographic images recorded by the pilots was magnitude -4 for a "large object". The pilots' visual sightings shortly before the photographs were taken indicated magnitude -5 for "4 or 5 very luminous point source objects".

The geometry of this incident corresponds to that shown in Figure 2. Specifically, the satellites were close to the horizon and near the solar azimuth while the Sun was 30° below the horizon. Furthermore, the BRDF function described in the previous section can be applied because the satellites from launch 4-26 were Starlink version 1.5.

The magnitudes listed above must be adjusted to a distance of 1000 km to correspond with the data in Figure 4. The mean range from Table 1 of Buettner et al is 1608 km, which indicates that the satellites would be 1.0 mag brighter at 1000 km based on the inverse square law of light. So, the apparent mags of -4 and -5 after adjustment become -5 and -6.

The BRDF values in Figure 4 are normalized to magnitude 0.0 for 0° off-specular, that is, directly on the beam. They are also 5.0 magnitudes fainter than that for the 1000-km observed magnitude axis. So, the BRDF magnitude on-beam and adjusted to the observed magnitudes is -5.0 as shown in Figure 5.

The agreement between magnitude -5.0 on the specular beam and the distance-adjusted values of -5 and -6 would suggest that the aircraft were near that beam. However, there were 52 satellites in launch 4-26. If each was equally bright then the distance-adjusted brightness would be fainter by 4.3 magnitudes which changes -5 and -6 to -0.7 and -1.7. Those brightness values indicate small but non-zero off-specular angles.

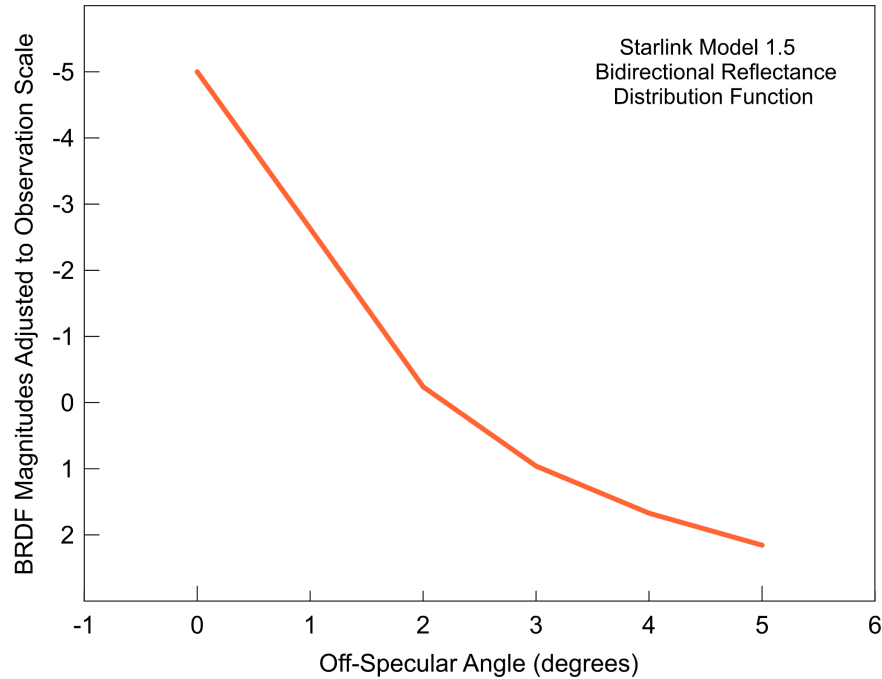

*Figure 5. BRDF function to 0º off-specular angle and adjusted to fit the observed magnitude scale.*

The exact value of the off-beam angle is difficult to determine for several reasons. (1) The observed magnitudes are only known to integer precision. (2) There are various numbers of objects studied in the image and the pilot visual sightings. (3) The Sun-satellite-observer geometry changed significantly during the incident. (4) It is unlikely that all the Starlink satellites were already in exactly the same attitude during their first day in orbit. (5) The solar arrays and other spacecraft components also contributed to the brightness. While the precise value of the off-beam angle is not known exactly, the analysis reported above suggests that it was about 2º or less when the individual bright objects were observed. The BRDF considerations discussed in Section 2 indicate that a much larger value would not account for the great brightness.

In summary, specular reflection from the nadir side of the Starlink spacecrafts' chassis can account for the brightness recorded by the pilots. This finding supports the conclusion of Buettner et al (2024) that the Unidentified Aerial Phenomena were actually Starlink satellites.

Acknowledgments

Thanks are due to Douglas Buettner for his review of the manuscript. The Heavens-Above.com web-site was used to plan the ground-based observations of flares. We appreciate the internal review of this paper by the IAU - Centre for the Protection of Dark and Quiet Skies from Satellite Constellation Interference which helped to improve it.